# Generalization of second order quasi-phase matching in whispering gallery mode resonators using Berry Phase

Alejandro Lorenzo-Ruiz[1,] Yoan Léger[1]
[1]Univ Rennes, CNRS, FOTON-UMR6082,F-35000 Rennes, France
e-mail address: yoan.leger@insa-rennes.fr

**Abstract:**
Second order nonlinearities in whispering gallery mode resonators are highly investigated for their many applications such as wavelength converters, entangled photon sources and generation of frequency combs. In such systems, depending on the material under scrutiny, the derivation of quasi-phase matching equations can lead to the appearance of additional quanta in the selection rule on the azimuthal confinement order. Here, we demonstrate that these additional quanta show up due to the Berry phase experienced by the transverse spin angular momentum components of the whispering gallery modes during their circulation within the resonator. We first detail the case of Zinc-blende materials and then generalize this theory to other crystal symmetries relevant for integrated photonics.

Second order nonlinear phenomena in integrated photonic devices are highly valued due to their potential use as frequency converters for on-chip optical interconnects,[1,2] in entangled photon sources for quantum processing[3–5] and even in the generation of supercontinuum.[6,7] The eligibility of many different materials for the realization of these devices is currently explored. We can cite in particular the integrated lithium niobate (LN) platform[5,8] as well as the III-V semiconductor one with experimental demonstrations using arsenides,[9–11] phosphides[12,13] or nitrides.[14–16] At the same time, planar whispering gallery mode (WGM) resonators seem to have established themselves as reference systems for nonlinear photonics on chip,[17,18] due to the convenience of their integration within photonic integrated circuits, in the form of microdisks, microrings or racetracks.

When nonlinear 2$^{nd}$ order phenomena are investigated in axisymmetric resonators, the conservation of photons momentum, the so-called phase matching condition, appears in the form of a selection rule on the azimuthal confinement orders of the input WGMs ($m_{in}$) and nonlinear products ($m_{prod}$) where additional quanta may appear depending on the crystal symmetry of the material under scrutiny. This was related to a "natural" poling of the crystal in these axisymmetrical geometries.[19–21] So Far, the most satisfying method to explain the occurrence of these additional quanta was the derivation of the Fourier components of the azimuthal dependence of the 2$^{nd}$ order nonlinear coefficient, providing momentum to the photons.[22] In planar z-cut LN-based WGM resonators where the extraordinary axis is parallel to the resonator symmetry axis, the currently observed trivial selection rule $\sum m_{prod} = \sum m_{in}$ can require artificial periodic poling to show efficient conversion.[23,24] On the contrary, "natural" quasi-phase matching (QPM) was reported in x-cut LN microdisks thanks to a 90° rotation of the crystal orientation which places the extraordinary axis of the crystal in the microdisk plane.[22] Second harmonic generation (SHG) was for example observed between a fundamental mode with $m_F = 111$ and a SH mode with $m_{SH} = 221$, demonstrating the loss of a single momentum quantum. The occurrence of such additional quanta in the momentum conservation rule had also been pointed out for III-V zinc-blende resonators.[25] This $\bar{4}$-QPM, named after the $\bar{4}3m$ point group symmetry of the crystal, has been predicted to require the condition:

$$\Delta m = \sum m_{prod} - \sum m_{in} = \pm 2 \qquad (1)$$

which has then been confirmed experimentally by many groups.[9,13] In contrast, the QPM selection rule of wurtzite GaN microdisks is simply $\Delta m = 0$.[16] In the calculation of the nonlinear conversion efficiency, each phase matching condition is associated to a weight which is specific to the signed value of the additional quantum.[20] Despite the directness of the Fourier approach, no clear physical meaning can be given to the expressions of these different weights. In addition, for resonators such as microdisks, the Fourier analysis approach requires some approximation on the effective radius of the modes. This points out the need of a new formalism for the derivation of 2nd order nonlinear processes in resonators with rotational symmetry.

Concurrently, a novel description of polarization properties of light in photonics has recently been introduced, based on the longitudinal electric field components showing up under certain conditions such as photonic confinement.[26,27] The later indeed enables the photons to feature transverse spin angular momentum (TSAM), a forbidden property for plane wave optics. In this framework, one can thus investigate the propagation of a photon carrying a spin (or helicity) perpendicular to the propagation plane. WGM resonators suit particularly well to this description since revolution symmetry is shared between the geometry of the system and the TSAM polarization basis as detailed in the following.

In this work, we use TSAM formalism to derive the field components of WGMs into a complete polarization basis respecting the revolution symmetry of the system and demonstrate that the TSAM components inherit a $\pm 2\pi$ Berry phase during the revolution of their transverse spin within the resonator. This additional topological phase is then considered in the derivation of second order nonlinear phenomena such as SHG which explains straightforwardly the origin of the diversity of natural QPM. We investigate in detail the case of $\bar{4}$-QPM in zinc-blende materials and finally generalize our approach to other nonlinear materials.

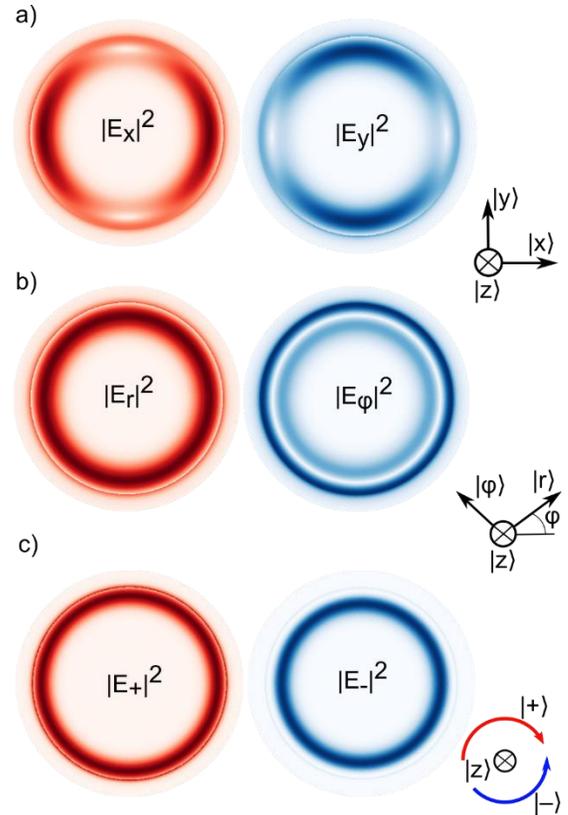

**Figure 1.** Squared magnitudes of the electric field planar components of a TE WGM using (a) a cartesian fixed basis, (b) a rotating polar basis and (c) the circular polarization basis along the resonator axis, unveiling the TSAM components of the WGM. The results are obtained from FEM simulations with a WGM at 1.9µm with azimuthal order m=18, planar and radial orders of 1, a refractive index of 3.04, radius of 3.4µm and 180nm of thickness.

In planar WGM resonators, the polarization eigenmodes of the system are commonly labeled as the ones of the slab waveguide: TE, with a main contribution of the electric field lays in the confinement plane and TM, which electric field pointing mainly outside of it (what we will call the z axis). For the sake of simplicity, we will make the approximation that TM modes feature only a non-zero z component of the electric field, so that only the TE modes will show a TSAM character. Going beyond this approximation, the full-vectorial model of NL processes even opens more possibilities to the present theory as suggested by Ciret *et al.*[28] While describing the mode profiles of the WGM, it straightforwardly comes that a frame following the symmetry of the system such as the rotating one ($|r\rangle, |\varphi\rangle, |z\rangle$) is more convenient than the fixed cartesian one ($|x\rangle, |y\rangle, |z\rangle$) as shown in Figure 1. Most theoretical works, from basic theory of WGMs to advanced derivation of NL quasi-phase matching have used this rotating frame.[20,21] The limitation of the later comes from the parametrized position of the basis unit vectors with the azimuthal angle $\varphi$. On the contrary, the fixed circular polarization (CP) frame allows a non-parametrized description of the WGMs and unveil their TSAM character. Let us define:

$$\begin{cases} |+\rangle = \frac{1}{\sqrt{2}}(|x\rangle + i|y\rangle) \\ |-\rangle = \frac{1}{\sqrt{2}}(|x\rangle - i|y\rangle) \end{cases} \quad (2)$$

the unit vectors of the CP frame describing the spin up and spin down polarization of photons. The $|z\rangle$ unit vector remains unchanged and can be seen as describing photons with spin 0. Within the disk, the electric field components of TE modes are well described in the rotating frame: [29]

$$\begin{cases} E_r = \frac{m}{r} C_m J_m(\tilde{k}_1 r) e^{im\varphi} \\ \quad = S_m(r) e^{im\varphi} \\ E_\varphi = \frac{i\tilde{k}_1}{2} C_m [J_{m-1}(\tilde{k}_1 r) - J_{m+1}(\tilde{k}_1 r)] e^{im\varphi} \\ \quad = i T_m(r) e^{im\varphi} \end{cases} \quad (3)$$

where m is the WGM azimuthal order, $C_m$ a constant, $J_i(x)$ Bessel functions of the first kind and $\tilde{k}_1$ the effective propagation constant of the WGM. The real-valued functions $S_m(r)$ and $T_m(r)$ are introduced for the sake of clarity. By using the transfer matrix $\mathcal{R}_z(\varphi)$ and the projections of eq.(2), one gets:

$$\begin{cases} E_+ = \frac{1}{\sqrt{2}}(S_m(r) + T_m(r)) e^{i(m-1)\varphi} \\ E_- = \frac{1}{\sqrt{2}}(S_m(r) - T_m(r)) e^{i(m+1)\varphi} \end{cases} \quad (4)$$

Opposite additional quanta on azimuthal dependence appears on both CP components, which come from the rotation operator involved in the rotating frame description and can also be written using the spin operator $s$ of the photon as $\mathcal{R}_z(\varphi) = \exp(-i\boldsymbol{\varphi} \cdot \boldsymbol{s})$.

This phenomenon is a direct consequence of the Berry phase experienced by each CP component during the revolution within the resonator as demonstrated by Chiao and coworkers:[30,31] During the round trip, the photon realizes a loop in parameter space $(k_x, k_y, k_z)$. The $k = 0$ point represents a topological monopole for photons and the photons should thus acquire a topological phase proportional to their spin number and to the solid angle subtended by the closed path with respect to the monopole. In Ref.[31], the Berry phase is investigated into an optical fiber helically wound around a cylinder and found to be $\Omega = 2\pi N(1 - \cos\theta)$ where N is the winding number of the helix and $\theta$ the local angle between the waveguide axis and the cylinder axis. Our case corresponds to an angle $\theta = \pi/2$ where a Berry phase $\Omega = 2\pi$ is acquired at each round trip, ensuring periodic boundary conditions in our resonant system. The azimuthal phase dependence of spin up and spin down photons of a TE WGM are thus different as presented in Figure 2 b) for m=18.

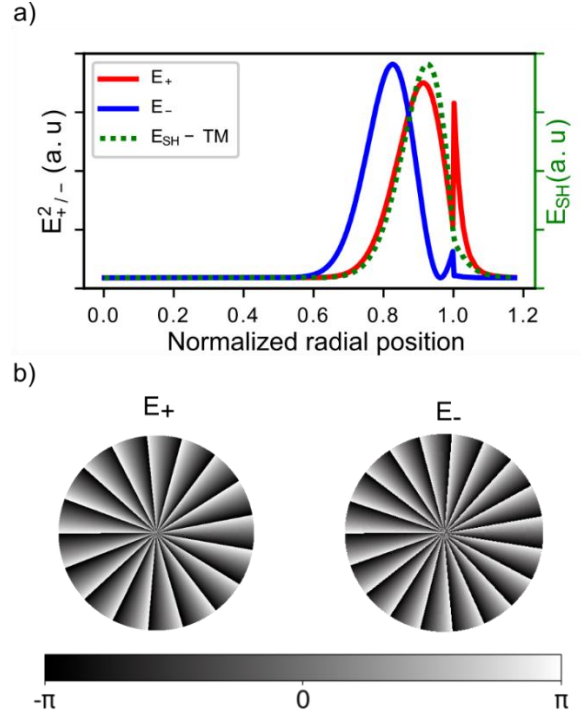

Figure 2. (a) radial profiles of the squared electric field components $E_+$ and $E_-$ of a fundamental TE mode together with $E_z$ radial profile of a TM SH mode for SHG in a $\bar{4}$-QPM WGM resonator (Parameters identical to Figure 1). (b) Phase distribution of the $E_+$ and $E_-$ components of the TE mode where opposite Berry phases shows up in the difference of $2\pi$ phase jumps (17 and 19 phase jumps respectively).

To unveil the effect of this Berry Phase of the CP field components on second order nonlinear phenomena, it is now necessary to project susceptibility tensors along the complete CP basis $(|+\rangle, |-\rangle, |z\rangle)$. Calling $\mathbf{Q}$ the transfer matrix from cartesian to CP basis, one can calculate the 2$^{nd}$ nonlinear polarization in the CP basis:

$$\boldsymbol{P}_{NL,CP} = \boldsymbol{Q} \cdot \chi^{(2)}_{x,y,z} \cdot (\boldsymbol{Q}^{-1} \cdot \boldsymbol{E}_{CP}) \otimes (\boldsymbol{Q}^{-1} \cdot \boldsymbol{E}_{CP}) \quad (5)$$

where $E_{CP}$ is the total input field and can contain contributions with different frequencies. Consequently the 2nd order susceptibility tensor can be defined as:

$$\chi^{(2)}_{CP} = Q \cdot \chi^{(2)}_{x,y,z} \cdot (Q^{-1} \otimes I_3) \cdot (I_3 \otimes Q^{-1}) \tag{6}$$

where $I_3$ is identity in the 3-dimensional vector space.

Let us first detail the case of SHG in a III-V zinc-blende microdisk resonator, featuring $\bar{4}$-QPM. We call $E_1$ the mode profile of the fundamental mode with azimuthal order $m_1$ and $E_2$ the mode profile of the SH one with azimuthal order $m_2$. Here, the only non-zero element of $\chi^{(2)}$ is $d_{14}$. Using eq. (6), the SH polarization can be expressed in the CP basis with contracted notations and reads as:

$$\begin{pmatrix} P_+ \\ P_- \\ P_z \end{pmatrix} = \epsilon\, d_{14} \begin{bmatrix} 0 & 0 & 0 & -i & 0 & 0 \\ 0 & 0 & 0 & 0 & i & 0 \\ i & -i & 0 & 0 & 0 & 0 \end{bmatrix} \begin{pmatrix} E_+ E_+ \\ E_- E_- \\ E_z E_z \\ 2E_- E_z \\ 2E_+ E_z \\ 2E_+ E_- \end{pmatrix} \tag{7}$$

We find in particular that a TM-polarized SH field can be generated from the nonlinear polarization $P_z$ and results from the square of either the spin up or the spin down field component of the fundamental mode, $E_+^2$ or $E_-^2$. The derivation of the nonlinear coupling coefficient for SHG $\beta_2$ as described for example by Rodriguez et al.[32] and which enters into account for the calculation of the conversion efficiency thus becomes:

$$\begin{aligned}
\beta_2 &= \frac{1}{4} \frac{\int d^3x \sum_{ijk} \epsilon \chi^{(2)}_{ijk} E^*_{2i} E_{1j} E_{1k}}{\left[\int d^3x\, \epsilon |E_1|^2\right]\left[\int d^3x\, \epsilon |E_2|^2\right]^{\frac{1}{2}}}, \quad i,j,k = x,y,z \\
&= \frac{i\, d_{14}}{4} \frac{\int d^3x\, \epsilon\, E^*_{2z}(E_{1+}^2 - E_{1-}^2)}{\left[\int d^3x\, \epsilon |E_1|^2\right]\left[\int d^3x\, \epsilon |E_2|^2\right]^{\frac{1}{2}}}
\end{aligned} \tag{8}$$

| z-cut LiNbO3 – Trigonal *3m* | x-cut LiNbO3 – Trigonal *3m* | GaN– Wurtzite P6₃mc | KNbO3 – *Amm2* |
|---|---|---|---|
| $2\epsilon_0 \begin{bmatrix} 0 & a & 0 & 0 & b & 0 \\ a^* & 0 & 0 & b & 0 & 0 \\ 0 & 0 & c & 0 & 0 & b \end{bmatrix}$ | $2\epsilon_0 \begin{bmatrix} a & b & c & 0 & 0 & a^* \\ b^* & a^* & c^* & 0 & 0 & a \\ 0 & 0 & 0 & c & c^* & 0 \end{bmatrix}$ | $2\epsilon_0 \begin{bmatrix} 0 & 0 & 0 & 0 & b & 0 \\ 0 & 0 & 0 & b & 0 & 0 \\ 0 & 0 & a & 0 & 0 & b \end{bmatrix}$ | $2\epsilon_0 \begin{bmatrix} 0 & 0 & 0 & a & b & 0 \\ 0 & 0 & 0 & b & a & 0 \\ a & a & c & 0 & 0 & b \end{bmatrix}$ |
| $d_{15} = d_{24}, d_{22} = -d_{16}$, $a = i\sqrt{2}d_{22}, b = d_{15}, c = d_{33}$ | $a = \sqrt{2}(-d_{33} + d_{15} + id_{22})/4$, $b = \sqrt{2}(-d_{33} - 3d_{15} + id_{22})/4$, $c = -\sqrt{2}(d_{15} - id_{22})/2$ | $d_{15} = d_{24}$, $a = d_{33}, b = d_{15}$ | $a = \frac{1}{2}(d_{15} - d_{24}), b = \frac{1}{2}(d_{15} + d_{24})$, $c = d_{33}$ |

| α-Quartz– *32* | BTM – Monoclinic *2* | Additional quanta to Δm |
|---|---|---|
| $2\epsilon_0 \begin{bmatrix} a & b & 0 & 0 & 0 & 0 \\ b & a & 0 & 0 & 0 & 0 \\ 0 & 0 & 0 & 0 & 0 & 0 \end{bmatrix}$ | $2\epsilon_0 \begin{bmatrix} a & b & c & d & 0 & a^* \\ b^* & a^* & c^* & 0 & d^* & a \\ d^* & d & 0 & c & c^* & 0 \end{bmatrix}$ | $\begin{bmatrix} +1 & -3 & -1 & -2 & 0 & -1 \\ +3 & -1 & +1 & 0 & +2 & +1 \\ +2 & -2 & 0 & -1 & +1 & 0 \end{bmatrix}$ |
| $a = \sqrt{2}(d_{11} + d_{12})/4$, $b = \sqrt{2}(d_{11} - 3d_{12})/4$ | $a = i\sqrt{2}(d_{16} + d_{22})/4$, $b = i\sqrt{2}(-3d_{16} + d_{22})/4$, $c = -id_{23}/\sqrt{2}$, $d = -id_{14}$ | |

**Table 1.** Contracted 2nd order susceptibility tensors expressed in the CP basis for different point groups symmetries and related materials. Values reported in pm/V. The bottom right matrix provides the list of additional quanta induced by the Berry phase mismatch in the azimuthal order selection rule for each conversion process, in the same contracted notation as equation 7. A general form of the tensor in the CP basis is provided in SI.

Inserting eq.(4) into eq. (8) allows to explicit the azimuthal dependence in the overlap integral. which leads to the $\bar{4}$-QPM conditions to be fulfilled for efficient SHG: $\Delta m \pm 2 = 0$, similar to eq. (1). The coupling coefficients $K_\pm$ introduced by Kuo in Ref.[21] can thus simply be attributed to the overlap integral of the SH mode with either the spin up or spin down components of the fundamental field as illustrated in Figure 2a. A discussion on the optimization of these overlaps is provided as supporting information as well as the whole derivation of nonlinear coupling coefficients and conversion efficiency.

We now generalize this approach to materials of interest for integrated nonlinear photonics. Table 1 provides the list of the additional quanta to be introduced in the azimuthal selection rule for each component association in processes such as SHG or sum frequency generation (SFG). The case of DFG requires a careful handling of conjugates in the CP basis which is discussed in the SI. Table 1 also gives a description of NL susceptibility tensors in the contracted CP basis for different point-group symmetries, illustrated with material cases relevant for integrated photonics. First it appears that each non-zero element of the tensor in the CP basis is now associated with a single QPM condition. At the 2$^{nd}$ order, up to seven QPM conditions can coexist, from $\Delta m = -3$ to $+3$, depending on the material symmetry. While for zinc-blende materials only $\pm 2$ conditions exist as already reported, the situation is different for LN. For z-cut LN, a natural QPM condition with $\Delta m = \pm 3$ in between two TE-polarized fields could thus be used to balance the material dispersion. However, it should feature a much smaller efficiency compared to the use of the $zzz$ tensor element both because of the value of the element itself and because of the possibly limited overlap between the $E_{+(-)}$ component of the fundamental field and the $E_{-(+)}$ component of the SH field. Note that the z-cut LN tensor does not present any QPM condition with $\Delta m = \pm 1$ as experimentally observed in Ref.[22]. In that case, prior to the CP basis projection, the LN tensor should be rotated by 90° to account for the x-cut. Notably, the resulting tensor shows that different SHG processes with $\Delta m = \pm 1$ are expected either with copolarized (TE) fundamental and SH fields[22] (elements $a$) or with cross polarized modes (element $c$).[19,33,34]

As a conclusion, we used the TSAM description of light in planar WGM resonators to explicit the Berry phase experienced by the spin up and spin down components of the field during their revolution into the resonator. This Berry phase is found to be at the origin of the additional quanta appearing in the quasi-phase matching conditions of nonlinear 2$^{nd}$ order processes in these devices for many different materials. This description does not only allow for a straightforward assessment of QPM conditions in WGM resonators; it also opens new routes for the design of more complex nonlinear processes in such integrated photonic devices.

### Acknowledgements
We acknowledge Yannick Dumeige and François Léo for fruitful discussions on nonlinear process theory in photonic devices.

This work is funded by the French national research agency through the project ORPHEUS ANR-17-CE24-0019-01 and Région Bretagne for the funding of predoctoral contract.

# Supporting information:

## A. General form of the 2nd order nonlinear tensor in the circular polarization basis and code

The general form of the nonlinear tensor can be written in the circular polarization basis with keeping the $d_{ij}$ parameters known from the usual cartesian representation of the tensor:

$$\begin{pmatrix} P_+ \\ P_- \\ P_z \end{pmatrix} = \chi^{(2)}_{CP} \begin{pmatrix} E_+ E_+ \\ E_+ E_- \\ E_z E_z \\ E_- E_+ \\ E_- E_- \\ E_- E_z \\ E_z E_+ \\ E_z E_- \\ E_z E_z \end{pmatrix} ; \qquad (S1)$$

$$\chi^{(2)}_{CP} = \epsilon_0 \begin{bmatrix} \frac{\sqrt{2}(d_{11}+d_{12}+id_{16}+id_{22})}{4} & \frac{\sqrt{2}(d_{11}+d_{12}-id_{16}-id_{22})}{4} & \frac{d_{15}}{2}+\frac{d_{24}}{2} & \frac{\sqrt{2}(d_{11}+d_{12}-id_{16}-id_{22})}{4} & \frac{\sqrt{2}(d_{11}-3d_{12}-3id_{16}+id_{22})}{4} & -id_{14}+\frac{d_{15}}{2}-\frac{d_{24}}{2} & \frac{d_{15}}{2}+\frac{d_{24}}{2} & -id_{14}+\frac{d_{15}}{2}-\frac{d_{24}}{2} & \frac{\sqrt{2}(d_{13}-id_{23})}{2} \\ \frac{\sqrt{2}(d_{11}-3d_{12}+3id_{16}-id_{22})}{4} & \frac{\sqrt{2}(d_{11}+d_{12}+id_{16}+id_{22})}{4} & id_{14}+\frac{d_{15}}{2}-\frac{d_{24}}{2} & \frac{\sqrt{2}(d_{11}+d_{12}+id_{16}+id_{22})}{4} & \frac{\sqrt{2}(d_{11}+d_{12}-id_{16}-id_{22})}{4} & \frac{d_{15}}{2}+\frac{d_{24}}{2} & id_{14}+\frac{d_{15}}{2}-\frac{d_{24}}{2} & \frac{d_{15}}{2}+\frac{d_{24}}{2} & \frac{\sqrt{2}(d_{13}+id_{23})}{2} \\ id_{14}+\frac{d_{15}}{2}-\frac{d_{24}}{2} & \frac{d_{15}}{2}+\frac{d_{24}}{2} & \frac{\sqrt{2}(d_{13}+id_{23})}{2} & \frac{d_{15}}{2}+\frac{d_{24}}{2} & -id_{14}+\frac{d_{15}}{2}-\frac{d_{24}}{2} & \frac{\sqrt{2}(d_{13}-id_{23})}{2} & \frac{\sqrt{2}(d_{13}+id_{23})}{2} & \frac{\sqrt{2}(d_{13}-id_{23})}{2} & d_{33} \end{bmatrix}$$

(S2)

In contracted notations, one gets :

$$\begin{pmatrix} P_+ \\ P_- \\ P_z \end{pmatrix} = \epsilon_0 \begin{bmatrix} \frac{\sqrt{2}(d_{11}+d_{12}+id_{16}+id_{22})}{4} & \frac{\sqrt{2}(d_{11}-3d_{12}-3id_{16}+id_{22})}{4} & \frac{\sqrt{2}(d_{13}-id_{23})}{2} & -id_{14}+\frac{d_{15}}{2}-\frac{d_{24}}{2} & \frac{d_{15}}{2}+\frac{d_{24}}{2} & \frac{\sqrt{2}(d_{11}+d_{12}-id_{16}-id_{22})}{4} \\ \frac{\sqrt{2}(d_{11}-3d_{12}+3id_{16}-id_{22})}{4} & \frac{\sqrt{2}(d_{11}+d_{12}-id_{16}-id_{22})}{4} & \frac{\sqrt{2}(d_{13}+id_{23})}{2} & \frac{d_{15}}{2}+\frac{d_{24}}{2} & id_{14}+\frac{d_{15}}{2}-\frac{d_{24}}{2} & \frac{\sqrt{2}(d_{11}+d_{12}+id_{16}+id_{22})}{4} \\ id_{14}+\frac{d_{15}}{2}-\frac{d_{24}}{2} & -id_{14}+\frac{d_{15}}{2}-\frac{d_{24}}{2} & d_{33} & \frac{\sqrt{2}(d_{13}-id_{23})}{2} & \frac{\sqrt{2}(d_{13}+id_{23})}{2} & \frac{d_{15}}{2}+\frac{d_{24}}{2} \end{bmatrix} \begin{pmatrix} E_+ E_+ \\ E_- E_- \\ E_z E_z \\ 2 E_- E_z \\ 2 E_+ E_z \\ 2 E_+ E_- \end{pmatrix}$$

(S3)

The code used to compute all these terms is provided below. To run the code, the installation of the sympy package for Python is needed.[S1] A version of the code is also available at https://github.com/Alex-l-r/QPM_BerryPhase_WGM.

```
from sympy.matrices import Matrix, eye, zeros, ones, diag, GramSchmidt
from sympy import symbols, pprint, sqrt, I, init_printing, simplify
from sympy.physics.quantum import TensorProduct

# Definition of the whole tensor in a x,y,z basis
d11, d12, d13, d14, d15, d16, d22, d23, d24, d33 = symbols("d11 d12 d13 d14 d15 d16 d22 d23 d24 d33")
susc = Matrix([[d11, d16, d15, d16, d12, d14, d15, d14, d13], [d16, d12, d14, d12, d22, d24, d14, d24, d23], [d15, d14, d13, d14, d24, d23, d13, d23, d33]])

# Definition of the transfer matrices for circular polarization basis
# Jones matrix, consistent with Rothberg lecture Washington university
rot = 1/sqrt(2)*Matrix([[1, -I], [1, I]])
```

```
rotm = 1/sqrt(2)*Matrix([[1, 1], [I, -I]])

#Full 3D polarization basis, including |z> polarization state
rot3 = zeros(3, 3)
rotm3 = zeros(3, 3)
rot3[0:2, 0:2] = rot
rotm3[0:2, 0:2] = rotm
rot3[2, 2] = 1
rotm3[2, 2] = 1

# Tensorial product towards the 3x3=9 basis
ROTmA = TensorProduct(rotm3, eye(3, 3))
ROTmB = TensorProduct(eye(3, 3), rotm3)

# Transformation
susc_CP = rot3*susc*ROTmA*ROTmB

# Compact form of the tensor
DCP = zeros(3, 6)
DCP[:, 0] = susc_CP[:, 0]
DCP[:, 1] = susc_CP[:, 4]
DCP[:, 2] = susc_CP[:, 8]
DCP[:, 3] = (susc_CP[:, 5]+susc_CP[:, 7])/2
DCP[:, 4] = (susc_CP[:, 2]+susc_CP[:, 6])/2
DCP[:, 5] = (susc_CP[:, 1]+susc_CP[:, 3])/2
DCP_simp = simplify(DCP)
```

## B. Field distribution of higher order fundamental WGMs in circular basis.

In order to compare the convenience of the use of higher radial order modes, we present here the radial profiles of the squared $E_{+/-}$ components of the first three radial orders for the fundamental modes together with profiles of the $E_z$ component of TM modes at the SH wavelength. This nonlinear conversion scheme applies to $\bar{4}$−QPM. The Figure S1 shows the utility of the CP-basis description, not only for explaining the physics beyond the QPM in circular microresonators but to optimize and design highly efficient devices.

Here the WGMs shown are calculated to satisfy the Δm=+2 condition, further calculations show that small changes in the azimuthal number barely modifies the distribution of the electric fields so we can use Fig.S1 to comment Δm=-2 cases too. It is important to remark that for real optimization of the SHG a careful and detailed optimization of the geometry of the resonator should also be made to obtain resonances for both WGMs.

For fundamental radial order equal to one, it is clear that for Δm=+2 ($E_-$ component), choosing a SH mode of radial order 2 gives a very good overlap. In the case where a process with $r_{SH}$=1 is wanted, using a microring instead of a microdisk will push the fundamental $E_-$ profile towards the external edge, improving the overlap with the SH $r_{SH}$=1 mode and thus the efficiency.

In contrast, the Δm=-2 QPM condition ($E_+$ component) should lead to maximal overlap with radial orders equals to 1, but it is made impossible in small disks due to chromatic dispersion.[S2] Looking at a fundamental radial order of 2, the combination with the SH WGM of r=3 gives a good overlap. In this situation, the chromatic dispersion might be compensated for some materials.

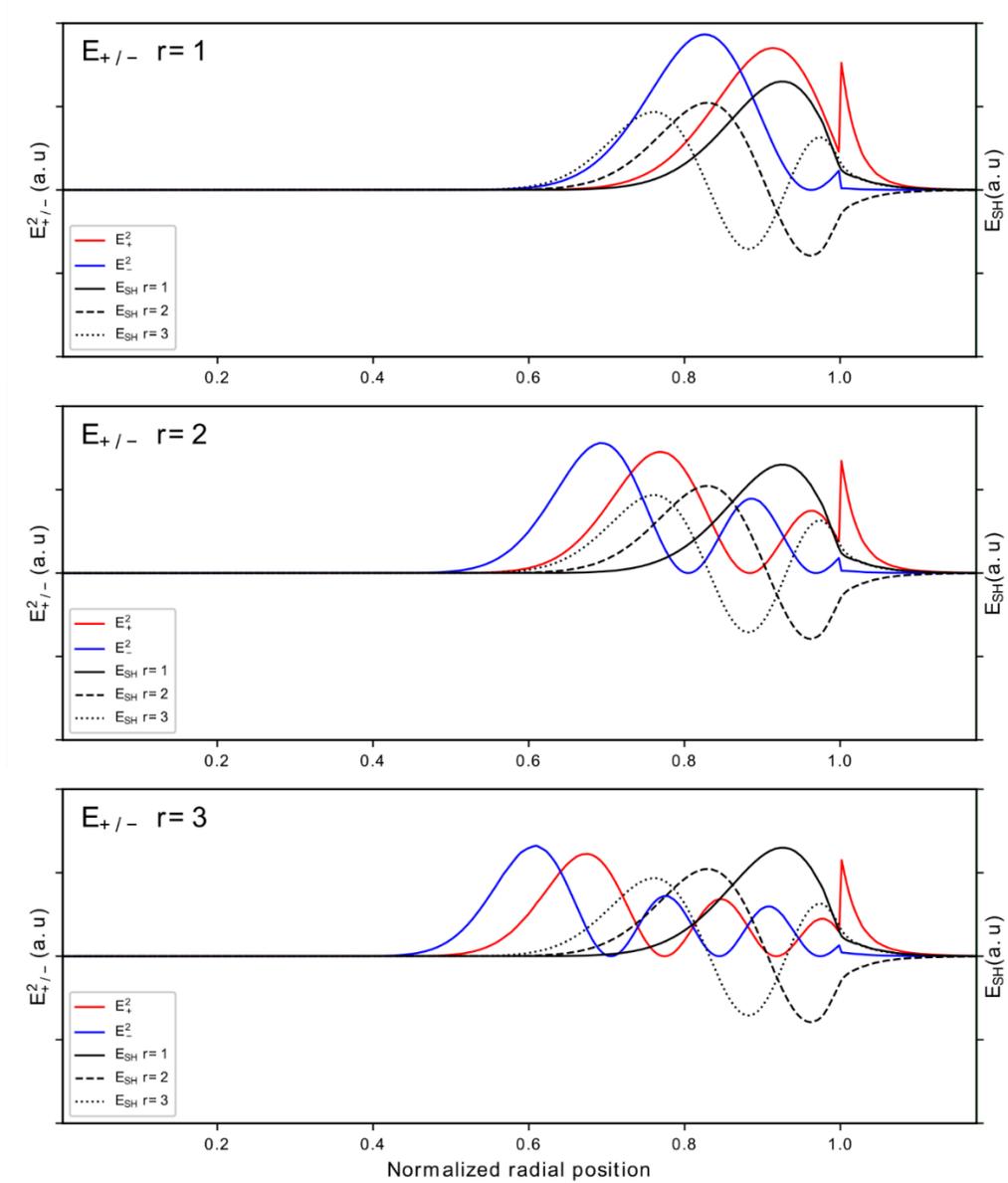

Figure S1: Representation of the radial profiles of $E_{+/-}$ components for the first three radial orders at fundamental wavelength, compared to the $E_z$ components of the first three radial orders of SH TM-polarized WGMs. The considered azimuthal orders are $m_F=18$ $m_{SH}=38$.

### C. Derivation of nonlinear coupling coefficients for time-dependent differential equations of 2nd order nonlinear processes.

Rodriguez et al.[S3] introduce the nonlinear differential equations describing second harmonic generation (SHG) in a doubly resonant cavity as:

$$\begin{cases} \frac{da_1}{dt} = \left(i\omega_1 - \frac{1}{\tau_1}\right)a_1 - i\omega_1\beta_1 a_1^* a_2 + \sqrt{\frac{2}{\tau_{c,1}}} F_1, \\ \frac{da_2}{dt} = \left(i\omega_2 - \frac{1}{\tau_2}\right)a_2 - i\omega_2\beta_2 a_1^2, \end{cases} \qquad (S4)$$

where $a_{1(2)}$ are the field envelopes of the fundamental (SH) cavity modes, featuring frequencies $\omega_{1(2)}$ and loaded decay times $\tau_{1(2)}$ defined as $1/\tau_{1(2)} = 1/\tau_{i,1(2)} + 1/\tau_{c,1(2)}$ with the index i holding for the intrinsic decay time and the index c holding for the coupling one. The decay times are related to the quality factors by: $Q_i = \omega_i \tau_i / 2$. The driving laser is modeled by $F_1$ and $\beta_1$ and $\beta_2$ are the nonlinear coupling coefficients:

$$\begin{cases} \beta_1 = \frac{1}{4} \frac{\int d^3x \sum_{ijk} \varepsilon \chi^{(2)}_{ijk} \left[E^*_{1i}\left(E_{2j}E^*_{1k} + E^*_{1j}E_{2k}\right)\right]}{\left[\int d^3x\, \varepsilon |E_1|^2\right]\left[\int d^3x\, \varepsilon |E_2|^2\right]^{\frac{1}{2}}} \\ \beta_2 = \frac{1}{4} \frac{\int d^3x \sum_{ijk} \varepsilon \chi^{(2)}_{ijk} E^*_{2i}E_{1j}E_{1k}}{\left[\int d^3x\, \varepsilon |E_1|^2\right]\left[\int d^3x\, \varepsilon |E_2|^2\right]^{\frac{1}{2}}} \end{cases} , \quad i,j,k = x, y, z \qquad (S5)$$

In the case of $\bar{4}$-QPM in WGM resonator and with the approximation that the only non-zero E field component of TM modes is along the WGM resonator axis so that only TE modes feature in-plane electric field components, SHG can only be obtained from a TE fundamental mode to a TM SH mode. The only field components to be considered are thus $E_{2z}$, $E_{1x}$, and $E_{1y}$. The transfer of the susceptibility tensor in the CP basis is quite straightforward for the calculation of $\beta_2$ since $2d_{14}E_{2z}^* E_{1x} E_{1y} = id_{14}E_{2z}^*(E_{1+}^2 - E_{1-}^2)$. However, the calculation of $\beta_1$ requires to extract the DFG process from the mixing of the real solution of the field: $E = \text{Re}(E_1 + E_2) = \frac{1}{2}(E_1 + E_2 + E_1^* + E_2^*)$. In the CP basis which feature complex coefficients, conjugation should be handled carefully:

$$\text{Re}(E) = \frac{1}{2}\begin{pmatrix} E_x + E_x^* \\ E_y + E_y^* \\ E_z + E_z^* \end{pmatrix}_{x,y,z} = \frac{1}{2\sqrt{2}}\begin{pmatrix} (E_x + E_x^*) - i(E_y + E_y^*) \\ (E_x + E_x^*) + i(E_y + E_y^*) \\ E_z + E_z^* \end{pmatrix}_{+,-,z} = \frac{1}{2\sqrt{2}}\begin{pmatrix} E_+ + E_-^* \\ E_- + E_+^* \\ E_z + E_z^* \end{pmatrix}_{+,-,z} \qquad (S6)$$

Taking into account this specific projection onto the CP basis unit vectors allows to restrict the sum on i,j, and k of the $\beta_1$ components to $-id_{14}E_{2z}(E_{1+}^{*2} - E_{1-}^{*2})$ so that $\beta_1$ and $\beta_2$ are complex conjugates. Using real solutions for $E_{2z}$, $H_{1z}$,[S4] and thus $E_{1+}$ and $E_{1-}$ as demonstrated in eq.(4) of the main text, and using the rotating frame for the sum over the microdisk volume, it comes that:

$$\beta_1 = \beta_2^* = \frac{i\pi}{2} \frac{\int r\,dr\,dz\, \varepsilon\, d_{14} E_{2z}(E_{1+}^2 - E_{1-}^2)}{\left[\int d^3x\, \varepsilon |E_1|^2\right]\left[\int d^3x\, \varepsilon |E_2|^2\right]^{\frac{1}{2}}} \qquad (S7)$$

The output SH field can be calculated from the intracavity field of (S4):

$$s_2 = \sqrt{\frac{2}{\tau_{c,2}}} a_2, \qquad (S8)$$

So that $P_{out} = |s_2|^2$ is the output SH power.

Solving the system of equations (S4) in the hypothesis of the non-depletion of the pump $a_1$ leads to:

$$P_{out} = |s_2|^2 = 2^6 \frac{\omega_2}{\omega_1^2} \left[\frac{Q_2^2}{Q_{c,2}}\right]\left[\frac{Q_1^2}{Q_{c,1}}\right]^2 \beta_2^2 P_L^2 = \eta\, P_L^2 \qquad (S9)$$

Where $P_L = |F_1|^2$ is the pump laser power. Eq.(S9) is consistent with the formulation proposed by Andronico *et al.* when using critical coupling condition.[S5]